\documentclass[a4paper,11pt]{article}
\pdfoutput=1 

\usepackage{jcappub} 

\usepackage[T1]{fontenc} 

\title{Probing sterile neutrino dark matter in the PTOLEMY-like experiment}

\author[a]{Ki-Young Choi,}
\author[a]{Erdenebulgan Lkhagvadorj}
\author[a]{and Seong Moon Yoo}

\affiliation[a]{Department of Physics, Sungkyunkwan University, 16419 Korea}

\emailAdd{kiyoungchoi@skku.edu}
\emailAdd{bulgaa@skku.edu}
\emailAdd{castledoor@skku.edu}

\abstract{We study the prospect to detect the cosmic background of  sterile neutrinos in the tritium $\beta$-decay, such as the PTOLEMY-like experiments. The sterile neutrino with mass between 1 eV - 10 keV may contribute to the local density as  warm or cold DM component. In this study, we investigate the possibility for searching them in the models with different production in the early Universe, without assuming  sterile neutrino as full dark matter component.
In these models, especially with low-reheating temperature and late-time phase transition, the capture rate per year can be greatly enhanced to be $\mathcal O(10)$ around the mass range $10\:-\:100\ \text{eV}$ without violating other astrophysical and cosmological observations.}

\abstract{We study the prospect to detect the cosmic background of  sterile neutrinos in the tritium $\beta$-decay, such as the PTOLEMY-like experiments. The sterile neutrino with mass between 1 eV - 10 keV may contribute to the local density as  warm or cold dark matter component. In this study, we investigate the possibility for searching them in the models with different production in the early Universe, without assuming  sterile neutrino as full dark matter component.
In these models, especially with low-reheating temperature and late-time phase transition, the capture rate per year can be greatly enhanced to be $\mathcal O(10)$ around the mass range $10\:-\:100\ \text{eV}$ without violating other astrophysical and cosmological observations.}

\begin{document}
\maketitle
\flushbottom

\section{Introduction}

The neutrino oscillation data observed in solar, atmospheric, reactor, and accelerator experiments requires non-vanishing mass~\cite{ParticleDataGroup:2022pth}, which is absent in the standard model (SM) of particle physics. Furthermore, it seems impossible that any particle in the SM can serve as a candidate for dark matter (DM) to explain the missing mass in the Universe. One of the simplest ways to explain both problems is to introduce right-handed neutrinos or sterile neutrinos~\cite{Drewes:2016upu,Dasgupta:2021ies}.  

Light sterile neutrinos around the eV scale, which have mixing with the active neutrinos, may explain the anomalies in the neutrino oscillations~\cite{MINOS:2020iqj, DayaBay:2016lkk, Gariazzo:2018mwd, Bilenky:1996rw, Dentler:2017tkw, Dentler:2018sju,Hagstotz:2020ukm}, while keV mass range sterile neutrinos may contribute significantly to DM density~\cite{Dodelson:1993je,Kusenko:2009up,Boyarsky:2018tvu}. The first analytical estimation of the relic energy density of sterile neutrinos was made by Dodelson and Widrow~\cite{Dodelson:1993je}. They assumed a negligible lepton number asymmetry, and sterile neutrinos are produced by thermal scatterings induced via active-sterile neutrinos oscillation in the early Universe. For keV mass, sterile neutrinos may be considered as a warm dark matter (WDM) candidate. In the presence of large lepton asymmetry, on the other hand, sterile neutrinos could be produced resonantly as proposed by Shi and Fuller~\cite{Shi:1998km}, with a non-thermal spectrum.  

Therefore, sterile neutrinos could exist as thermal relics, similar to the cosmic microwave background (CMB) or cosmic neutrino background (C$\nu$B). Nonetheless, in contrast to the global relic density of sterile neutrinos, local density around Sun may increase due to the gravitational clustering effect. By employing a local clustering effect, obtained from interpolating the N-body and N-1-body simulation results~\cite{Anderhalden:2012qt,Ringwald:2004np,deSalas:2017wtt,Zhang:2017ljh,Mertsch:2019qjv}, we could calculate an enhanced number density for sterile neutrinos. A more detailed discussion on the interpolation is provided in Section \ref{capture}.

 The direct detection of C$\nu$B can be done in the capture of the electron neutrino on the radioactive $\beta$-decaying nuclei with resultant peak in the electron spectrum.
The PTOLEMY experiment proposes using 100g tritium as a target coated on graphene~\cite{PTOLEMY:2018jst,PTOLEMY:2019hkd,Betti:2018bjv,Apponi:2021hdu,PTOLEMY:2022ldz}. Its energy resolution is the order of the neutrino mass scale $\Delta \sim 0.15$ eV. In this experiment, it is expected that around 4 events for Dirac neutrinos and 8 events for Majorana neutrinos per year could be detected on this target. Although the direct detection of the C$\nu$B signal seems challenging with small neutrino masses, several phenomenological aspects and sensitivity estimates of PTOLEMY experiment have been done in the Literature~\cite{Long:2014zva,McKeen:2018xyz,Chacko:2018uke,Bondarenko:2020vta,Akita:2021hqn,Alvey:2021xmq,Hufnagel:2021pso,Dror:2019dib}.  

Since the sterile neutrino can have mixing with electron neutrino, the cosmic sterile neutrino background (C$\nu_s$B) also can be measured on the radioactive $\beta$-decaying nuclei. The resulting electron spectrum can show a peak which is distinguishable from the ordinary $\beta$-decay spectrum or from C$\nu$B. Probing cosmic sterile neutrino background was studied in~\cite{Liao:2010yx,Li:2010sn,Li:2010vy,Lasserre:2016eot}, with assuming that most of the DM was sterile neutrino with mass around keV scale. If sterile neutrinos compose 100\% of DM, the mass-mixing parameter space is constrained by astrophysical observations such as phase space bounds~\cite{Tremaine:1979we, Alvey:2020xsk}, Lyman-$\alpha$ forest~\cite{Viel:2005qj, Hooper:2022byl}, and X-ray emissions~\cite{Ng:2019gch, Roach:2019ctw}, with a viable mass between $1 \ {\rm keV} < m_s < 50  \ {\rm keV}$ and mixing $10^{-13} < |U_{e4}|^2<10^{-7}$.
 In that case, the number of event can be $\mathcal O(1)$ per year with 10 kg tritium or 10 ton Ru, which is quite difficult to detect.
 
In this paper, our primary contributions include studying the prospects for detecting C$\nu_s$B using the PTOLEMY-like experiments without assuming 100\% of DM with sterile neutrino, as well as exploring models with low-reheating temperature~\cite{Gelmini_2004,Hasegawa:2020ctq,Gelmini:2019wfp,Gelmini:2020duq,Gelmini:2019clw,Yaguna:2007wi,Moroi:2020has,Benso:2019jog} and late phase transition~\cite{Bezrukov_2017} that predict different relic abundance of the sterile neutrino DM. Within these models, the production of the sterile neutrino is suppressed compared to that in the standard Dodelson-Widrow mechanism, and thus a large mixing angle is needed to achieve the corresponding relic abundance. Consequently, we find that the larger mixing angles around $|U_{e4}|^2\sim 10^{-3}$ could enhance the capture rate of C$\nu_s$B up to $\mathcal O(10)$ per year for sterile neutrinos with masses in the 10 - 100 eV range without violating other astrophysical and cosmological observations.

The paper is organized as follows. In section \ref{ptolemy}, we focus on the PTOLEMY-like experiment and review the detection of the standard C$\nu$B. In section \ref{sterile}, we discuss the capture rate of cosmic relic sterile neutrinos which could be sub-dominant component of DM and clustered near the Earth. We introduce a new fitting function of clustering effect, so we could obtain the local number density of sterile neutrinos. In section \ref{constraints}, we review the present constraints on sterile neutrinos including oscillation experiments, $\beta$-decay experiments, X-ray telescopes, phase space bound, Lyman-$\alpha$ forest, big bang nucleosynthesis (BBN) and CMB bounds in the early universe. In section \ref{DW}, we calculate the number density of clustered sterile neutrinos and capture rate of them with clustering effect in the standard Dodelson-Widrow mechanism. In section \ref{lowT}, we discuss non-standard models based on low reheating temperature and late phase transition in the hidden sector.
 Finally, we present our conclusions and outlooks in section \ref{con}.

\section{PTOLEMY-like experiment}
\label{ptolemy}

In nature, the $\beta$-decay is a spontaneous process, which does not have energy barrier. For example, the tritium (${}^{3}\text{H}$) can decay with its half-life 12.32 years,
 to  Helium (${}^{3}\text{He}$), electron ($e^-$) and $i$-th mass eigenstate of anti-neutrino ($\bar{\nu}_i$),  $  {}^{3}\text{H}\rightarrow {}^{3}\text{He} + e^- + \bar{\nu}_i$. 
In its inverse $\beta$-decay, the neutrino can be captured by tritium, and produce ${}^{3}\text{He}$ and  $e^-$:
\begin{equation}
    \nu_i+{}^{3}\text{H}\rightarrow {}^{3}\text{He} + e^-.
\end{equation}
This process can happen even with non-relativistic neutrinos and their energy is converted to the kinetic energy of the electron in the final stage $K_e=E_e  -m_e$. This energy is displaced from the tritium $\beta$-decay endpoint $K_{\rm end}$~\cite{Long:2014zva}
\dis{
K_{\rm end} = \frac{ (m_{{ }^{3} \mathrm{H}}-m_e )^2 -(m_{{ }^{3} \mathrm{He}} + m_{\nu})^2}{2m_{{ }^{3} \mathrm{H}}}  .
}
For $m_{{ }^{3} \mathrm{H}}\simeq m_{{ }^{3} \mathrm{He}} \gg m_e \gg m_\nu$, the electron energy from  $C\nu B$ is 
\dis{
K_e^{ C\nu B} \simeq K_{\rm end}+2m_\nu \simeq K_{\rm end,0}+m_\nu,
}
where $K_{\rm end,0}\simeq 18.5988 \kev$ is $ K_{\rm end}$ for massless neutrino.

The PTOLEMY experiment~\cite{PTOLEMY:2018jst,PTOLEMY:2019hkd} has been proposed to probe the background neutrinos using the inverse $\beta$-decay by measuring precisely the energy spectrum of the final electrons. The energy resolution of $\Delta \sim 0.15$ eV is expected to be obtained with a 100 g sample of tritium. Moreover, in order to distinguish the signal of the neutrino capture from the tritium $\beta$-decay, the energy resolution $\Delta$ should be smaller than half of the  neutrino mass, $\Delta \lesssim m_\nu/2$.

Considering the energy resolution  $\Delta$, the electron spectrum is convoluted with a Gaussian envelope of FWHM  $\Delta=2\sqrt{2\ln 2} \sigma\simeq 2.35\sigma$, with $\sigma$ the standard deviation of the Gaussian. 
The observed spectrum after convolution~\cite{Long:2014zva}, 
\begin{equation}\label{electron spectrum}
\frac{d\Gamma }{dE_e} =\frac{1}{\sqrt{2\pi}\sigma }\int_{-\infty}^{+\infty}dE_e'\frac{d\tilde{\Gamma}}{dE_e'}(E_e')\exp\left[ - \frac{(E_e'-E_e)^2}{2\sigma^2}\right],
\end{equation}
where $\frac{d\tilde{\Gamma}}{dE_e'}(E_e')$ is the true spectrum of electron. This has two contributions from the $\beta$-decay and the cosmic neutrino contributions. The sterile neutrino background can give additional contribution to these, which will be the main subject in this paper.

The $\beta$-decay spectrum is given by
\begin{equation}
    \frac{d\tilde{\Gamma}_\beta}{dE_e'} (E_e')= \sum_{j=1}^{3}|U_{ej}|^2\frac{\bar{\sigma}}{\pi^2}H(E_e',m_{\nu_j}) N_T,
\end{equation}
where $U$ is the Pontecorvo-Maki-Nakagawa-Sakata (PMNS) matrix~\cite{Mohapatra_2007}, $\bar{\sigma}$ is the capture cross section of the electron neutrino defined in~\cite{Long:2014zva}. For non-relativistic limit
\begin{equation}
    \bar{\sigma} \simeq \sigma_e v_\nu \simeq 3.834\times 10^{-45} \, {\mathrm \cm}^2,
\end{equation}
and 
\begin{equation}
\left.H\left(E_{e}', m_{\nu_{j}}\right) = \frac{1-m_{e}^{2} /\left(E_{e}' m_{{ }^{3} \mathrm{H}}\right)}{\left(1-2 E_{e}' / m_{{ }^{\mathrm{H}}}+m_{e}^{2} / m_{3_{\mathrm{H}}}^{2}\right)^{2}} \sqrt{y\left(y+\frac{2 m_{\nu_{j}} m_{{ }^{3} \mathrm{He}}}{m_{{ }^{\mathrm{H}}}}\right.}\right)\left[y+\frac{m_{\nu_{j}}}{m_{{ }^{3} \mathrm{H}}}\left(m_{{ }^{3} \mathrm{He}}+m_{\nu_{j}}\right)\right].
\end{equation}
Here $y=m_e + K_{\rm end} - E_e'$ and $N_T=m_T/m_{{ }^{3} \mathrm{H}}$ is the approximate number of the nuclei in the sample.

The electron spectrum from the cosmic neutrino background is given by
\begin{equation}
    \frac{d\tilde{\Gamma}_{C\nu B}}{dE_e'} =\Gamma_{C\nu B}\, \delta[E_e' -(E_{\rm end}+2m_\nu)],
    \label{Espectrum_cnub}
\end{equation}
where for the non-relativistic neutrinos,  the rate $\Gamma_{C\nu B}$ is given by~\cite{Long:2014zva} for  Dirac neutrinos,
\dis{
\Gamma_{C\nu B}^{\rm D} = \bar{\sigma} n_0 N_T,
\label{rate_D}
}
and for  Majorana neutrinos,
\dis{
\Gamma_{C\nu B}^{\rm M} = 2\bar{\sigma} n_0 N_T,
\label{rate_M}
}
with  the cosmological average of the neutrino number density $n_0 =56\, {\rm cm}^{-3}$.
Here we assumed that clustering effects for the neutrino is negligible and also used the unitarity of the PMNS matrix $\sum_i |U_{ei}|^2=1$.

 \section{Capture rate of Sterile neutrino Dark Matter}\label{capture}
\label{sterile}

The sterile neutrinos are produced in the early Universe and comprise a component of hot, warm or cold dark matter.
Once the sterile neutrino exists as a background in our Milky Way, it can be captured by the tritium through mixing with the electron neutrino $U_{e4}$. This small mixing suppresses the capture rate of the sterile neutrino compared to the active ones and makes it more difficult to probe. However massive sterile neutrinos can cluster and enhance the local density around the Earth. 
 The relic density of the sterile neutrino also can be modified depending on the production models in the early Universe and the cosmological constraints can be relaxed. In this section, we summarise the capture rate of the sterile neutrino and the relevant constraints on them in the next section. 

The capture rate of the sterile neutrino can be obtained using the equations for the active neutrinos, except the mixing angle and the relic number density. The capture rate $\Gamma_{C\nu B}$ in~\eq{Espectrum_cnub} should be modified to the rate for Majorana sterile neutrino $\Gamma_{C\nu_{s} B}$ with
\dis{
   \Gamma_{C\nu_{s} B} = N_T |U_{e4}|^2\int dE_{\nu_4} \sigma_e v_{\nu_4} \frac{dn_{\nu_4}}{dE_{\nu_4}} \simeq N_T |U_{e4}|^2 \bar{\sigma} n_{s, \text{loc}},
}
where we assumed that $\sigma_e v_{\nu_4}$ is energy-independent for low velocity and approximates to be $\bar{\sigma}$  and  $n_{s,\text{loc}} =\int dE_{\nu_4}  \frac{dn_{\nu_4}}{dE_{\nu_4}}$ is the local number density of the sterile neutrino near the Earth.

For massive sterile neutrinos, their number density near the Sun in the Milky Way is larger than that of the cosmological relic density due to the gravitational clustering.
The clustering effect near the Earth is parameterized with a parameter $f_c$ given by
\dis{
n_{s,\text{loc}} = (1+f_c(m_s))n_{s}=n_s+n_{s,\text{cls}},
\label{nsloc}
}
where $n_s$ is the global number density before clustering. The number density due to the clustering should be smaller than the bound from the local phase space constraint, which will be discussed later in Section~\ref{phasespacebound}.

The clustering parameter $f_c$ may depend on the mass and momentum distribution of the sterile neutrino.
Once they are produced thermally in the early Universe, the distribution function can be the similar form as the thermally produced particles such as active neutrinos and  WDM~\cite{Ringwald:2004np,Anderhalden:2012qt,deSalas:2017wtt,Zhang:2017ljh,Mertsch:2019qjv}. To calculate the clustering effect properly, it is necessary to do N-body simulation, which is beyond in this work. Instead we used the results from N-body~\cite{Anderhalden:2012qt} and N-1-body~\cite{Ringwald:2004np, Zhang:2017ljh, deSalas:2017wtt} simulations which were performed under the NFW profile with baryonic contribution. We interpolated those results and found a fitting function of $f_c$ as follows:
\begin{equation}\label{eq:fitting}
f_c(m_s) = f_{c,\text{DM}} \left[1+\left(a \ \frac{\text{keV}}{m_s}\right)^b\right]^{-c/b},
\end{equation}
where we normalized the clustering factor of cold DM $f_{c,\text{DM}}\approx2.4 \times 10^5$ corresponding to the local DM desntiy $\rho_\text{DM,loc}=0.3 \text{ GeV/cm}^3$.
To fit the results, we employed the root mean squared logarithmic error as a cost function, and utilized the Adam optimization algorithm to minimize this cost function.
We found that the optimal values are $a=0.037$, $b=2.61$, and $c=2.3$.

\begin{figure}[tbp]
\centering
 \includegraphics[width=0.45\textwidth]{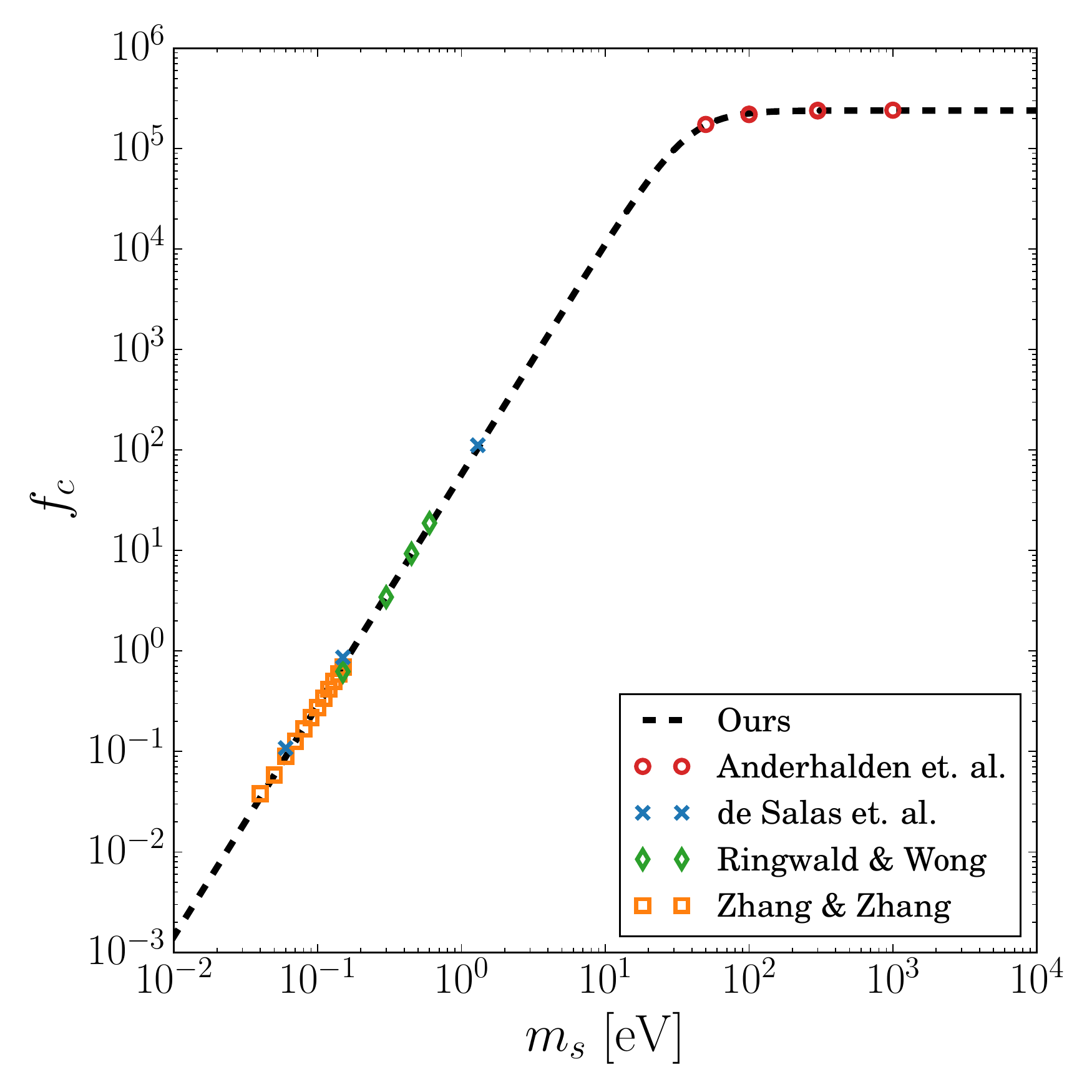}
\caption{The clustering of $f_c$ vs mass of sterile neutrino $m_s$ near the Sun in our Milky Way. The different simulation results are shown with red circle~\cite{Anderhalden:2012qt}, blue cross sign~\cite{deSalas:2017wtt},  green diamond~\cite{Ringwald:2004np}, and orange square~\cite{Zhang:2017ljh}, respectively. The optimal fitting function is shown with black dashes line, with parameters $a=0.037$, $b=2.61$, and $c=2.3$ according to Eq.~(\ref{eq:fitting}).
}
\label{fig:fitting}
\end{figure}

The figure~\ref{fig:fitting} shows the clustering effect of the sterile neutrino with a mass $m_s$ near the Sun.  The different simulation results are shown with red circle~\cite{Anderhalden:2012qt}, blue cross sign~\cite{deSalas:2017wtt},  green diamond~\cite{Ringwald:2004np}, and orange square~\cite{Zhang:2017ljh}, respectively.
Our optimal fitting function is shown with a dashed black line. However, we note that the clustering may change for different local DM density and the density profile of DM and baryonic matter.  The largest uncertainty comes from the local DM denstiy which is between  $0.3\ \text{GeV/cm}^3$ to around $0.7\ \text{GeV/cm}^3$~\cite{deSalas:2019pee}. Therefore, the clustering effect may enhance by around factor 2 from our optimal fitting function.

We use a parameter $\omega$ to denote the fraction of the energy density of the sterile neutrino in the total dark matter as follows:
\dis{
\begin{split}
    \omega_s &\equiv \frac{\rho_{s}}{\rho_{\rm DM}} = \frac{m_s n_{s}}{m_{\rm DM} n_{\rm DM}},\\
    \omega_{s,\rm{loc}}&\equiv\frac{\rho_{s,\text{loc}}}{\rho_{\rm DM,\text{loc}}} = \omega_s\frac{f_c}{f_{c,\DM}},
\end{split}
} where $\rho_s$ and $\rho_{s,\text{loc}}$ are the global and local energy density of sterile neutrinos, while $\rho_\text{DM}$ is the global energy density of DM.

\begin{figure}[tbp]
\begin{center}
 \includegraphics[width=0.45\textwidth]{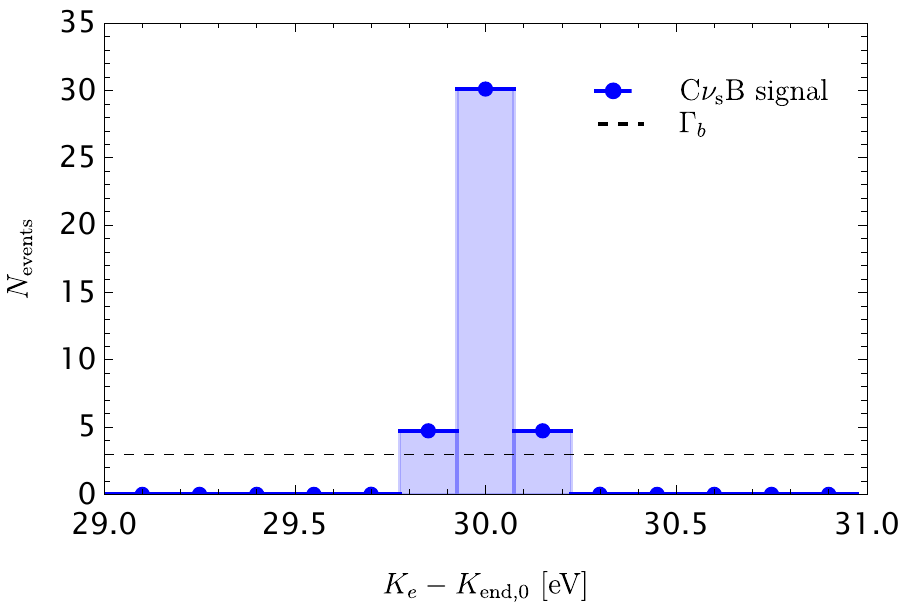}
\end{center}
\caption{Number of events of C$\nu_s$B per energy bin $\Delta E = 0.15 $ eV for one year of exposure time in the future PTOLEMY-like experiment with benchmark values with $m_s=30\ \rm{eV}$ and $|U_{e4}|^2=10^{-3}$. We used the local number density of the sterile neutrino $n_{s,\text{loc}}\simeq 5.4\times 10^5 \ \rm{cm}^{-3}$ and the expected background rate $\Gamma_b \simeq 10^{-7}$ Hz~\cite{PTOLEMY:2019hkd} per energy bin.
}
\label{fig:electronspectrum}
\end{figure}

In figure~\ref{fig:electronspectrum}, we show the number of events of C$\nu_s$B per energy bin with $\Delta E = 0.15 $ eV for one year of exposure time in the future PTOLEMY-like experiment. The number of events per energy within an energy bin centered at $E_k$ is calculated as~\cite{PTOLEMY:2019hkd, Alvey:2021xmq}, 
\dis{
N_{k} = t_{\rm yr} \int_{E_k - \Delta/2}^{E_k + \Delta/2} \frac{d\Gamma_{C\nu_s B}}{dE_e} dE_e,
}
where exposure time $t_{\rm yr}=1$ year and the size of energy bins equal to the detector resolution $\Delta = 150$ meV.
The dashed black line represents a fiducial PTOLEMY background rate $\Gamma_b \simeq 10^{-7}$ Hz~\cite{PTOLEMY:2019hkd} per energy bin. We obtain around 30 signal events per year with $m_s=30\ \rm{eV}$ and $|U_{e4}|^2=10^{-3}$ assuming $n_{s,\text{loc}}\simeq 5.4\times 10^5 \ \rm{cm}^{-3}$. 

\section{Constraints on Sterile neutrino}\label{constraints}
In this paper, we are mainly focused on the mixing of the sterile neutrino with the electron neutrinos to see the possibility for the sterile neutrino capture. This mixing can also affect the neutrino oscillation experiments, $\beta$-decay experiments, X-ray constraints, and cosmological observations. In this section, we summarize the possible constraints from these observations on the mixing of the sterile neutrinos.

\subsection{Oscillation experiments}
The neutrino oscillation experiments to measure  the appearance and disappearance of the neutrinos can constrain the mixings of the sterile neutrino to the active neutrinos.
The Daya Bay and Bugey-3 reactor experiments provide an upper limit of $\sin^22\theta_{14} \lesssim 0.06$ at 90\% C.L. around $\Delta m^2_{41}\approx1.75 \ \text{eV}^2$ \cite{DayaBay:2016lkk,MINOS:2020iqj}. In the short-baseline approximation \cite{Bilenky:1996rw}, this bound corresponds to $|U_{e4}|^2\lesssim0.015$. 

To explain the $\nu_e$ disappearance anomaly and the NEOS/Daya Bay and DANSS overlapped analysis, non-zero mixing is preferred at about $\sim$ 3$\sigma$ with best-fit point  of $m_{41}^2 \approx 1.3 \ \text{eV}^2 \ \text{and} \ |U_{e4}|^2 \approx 0.012$ \cite{Dentler:2018sju, Gariazzo:2018mwd, Dentler:2017tkw}.

\subsection{\texorpdfstring{$\beta$}{\texttwoinferior}-decay experiments}
The $\beta$-decay of tritium can produce sterile neutrinos through mixing, which leads to a distortion in the electron energy spectrum. The current constraints on the mixing parameter, established by the non-detection of such distortion, give $|U_{\mathrm{e}4}|^2\lesssim 10^{-2} - 10^{-3}$ for sterile neutrino masses ranging from 10 eV to 1 keV~\cite{Hiddemann:1995ce,Belesev:2013cba,Abdurashitov:2017kka,KATRIN:2022ith,KATRIN:2022spi}.

The future expectations for the PTOLEMY experiment with 100 grams of tritium would probe the mixing in the region $|U_{\mathrm{e}4}|^2 \sim 10^{-4} - 10^{-6}$ for sterile neutrino masses between 0.1 and a few eV~\cite{PTOLEMY:2019hkd}.

\subsection{X-ray telescope}

The dominant decay mode of the sterile neutrino is the decay into 3 active neutrinos mediated by Z-boson and the decay rate is given by~\cite{Dasgupta:2021ies}
\dis{
\Gamma (\nu_s \rightarrow \nu_\alpha + \nu_\beta+\bar{\nu}_\beta) \simeq \frac{2G_F^2m_s^5}{192\pi^3}\sum_{\alpha =e}^{\mu}|U_{\alpha s}|^2,
}
for $m_s \gg m_i$.
The corresponding lifetime is 
\dis{
\tau \simeq 1.44\times 10^{27}\, \textrm{s}\bfrac{1 \kev}{m_s}^5\frac{10^{-8}}{\sum|U_{\alpha s}|^2}.
}
This should be larger than the age of the Universe to be probed in the PTOLEMY-like experiments.

Apart from the Z-boson decay channel, one-loop decay channel through W-boson emits the photon, with a decay rate as given by~\cite{Dasgupta:2021ies}
\dis{
\Gamma(\nu_s\rightarrow \nu_a+\gamma) = 4.4\times 10^{-29} \, {\rm sec}^{-1} \bfrac{\sin^2 2 \theta_s}{10^{-8}} \bfrac{m_s}{1\kev}^5.
}
 The resulting photon could be observed in the X-ray telescope. Although several observations have been made to detect the X-ray light from sterile neutrino DM, the upper bound on the decay rate and the abundance of sterile neutrino DM is only given. In our study, we used the results from M31 by Chandra X-ray observatory~\cite{Horiuchi:2013noa} and the galactic bulge observation of NuSTAR~\cite{Ng:2019gch,Roach:2019ctw}. 

The conventional bound on the sterile neutrino mixing to the active neutrino assumes that the sterile neutrino explains whole DM. Since we don't assume that in this study, we reproduce the mixing with the corresponding relic density of the sterile neutrino. Therefore, the new bound can be written as
\dis{
|U_{e4}|^2_{\omega_s<1} = 
  \left( \frac{\Omega_{\rm DM, local}}{\Omega_\text{s,local}}\right )|U_{e4}|^2_{\omega_s=1}.
}

\subsection{Phase space bound}
\label{phasespacebound}
Identical fermionic particles cannot occupy the same quantum state. 
Therefore for a given escape velocity (or a momentum), there exists a maximum number density when the particles occupy the energy states form the lowest level~\cite{Tremaine:1979we,Hufnagel:2021pso}. This gives the bound on the relic density of fermionic DM.

For degenerate fermion up to the maximum momentum $p_{\rm max}$, the number density is bounded from the value with the distribution function $f_s=1$ and thus
\dis{
n_s =g_s \int_0^{p_{\rm max}}  f_s \frac{d^3 p }{(2\pi)^3} \lesssim  g_s \frac{p_{\rm max}^3}{6\pi^2},
}
with the degrees of freedom of $g_s$.
Near the Earth of the Milky,  $p_{\rm max} = m_s v_\text{esc}$ with the escape velocity $v_\text{esc} \simeq 550$km/s  for non-relativistic DM, the maximum number density of clustered sterile neutrino  is bounded as
\begin{equation}
    n_{s,\text{cls}} \leq g_s\frac{(m_s v_\text{esc})^3}{6\pi^2}.
    \label{phasespace}
\end{equation}

\subsection{Constraints from early Universe}
Even though the sterile neutrinos are non-relativistic in the present Universe, they can be relativistic in the early Universe, since their momentum redshifts with the expansion of the Universe. The relativistic component may affect BBN, CMB, and the small scale of the structure formation in the early Universe.

The extra relativistic component is usually parameterized by $\Delta N_{\text{eff}}$ from the relation 
\dis{
\rho_{s} = \Delta N_{\text{eff}}\,\frac{7}{8} \bfrac{4}{11}^{4/3}\rho_\gamma,
}
where $\rho_\gamma = \frac{\pi^2}{15}T_\gamma^4$. Below we use the constraint from Planck 2018, $N_{\text{eff}}=3.04+\Delta N_{\text{eff}} <3.29$ and $m^{\rm eff}_{\nu, \rm sterile}<0.65\, {\rm eV}$ for the mass $m_s < 10 \,{\rm eV}$~\cite{Planck:2018vyg}.

For the constraints from the Lyman-$\alpha$ forest observations, we use the constraints on the thermal warm dark matter (WDM)~\cite{Hooper:2022byl}, where the constraints are given on the parameter space of the WDM mass $m_w$ and the relic density $\Omega_w$,
\dis{
m_w \gtrsim m_w^{L-\alpha} \equiv 7.2 \kev \left( \frac{\Omega_w}{\Omega_{\rm DM}} -0.1\right).
\label{Lyman_WDM}
}
For the same relic density of the sterile neutrino as WDM, we can find the corresponding
mass of sterile neutrino which gives the same free-streaming scale, and then that is the lower bound for the given relic density from the Lyman-$\alpha$ forest observation~\cite{Viel:2005qj}.
Note that the free-streaming scale is determined by the background temperature $T_{NR}$ when the sterile neutrino becomes non-relativistic and $T_{NR}$  can be written in terms of its mass and the temperature of the sterile neutrino as
\dis{
T_{NR} = T_{s,NR} \bfrac{T_{NR}}{T_{s,NR}} \simeq \frac{m_s}{3} \bfrac{T}{T_{s}},
}
where $T_{s,NR}$ is the temperature of the sterile neutrino when it becomes non-relativistic, which is $T_{s,NR}=m_s/3$ for thermally produced case, and $T_{NR}$ is the temperature of the background plasma at the same time.
In the second equality we used that the temperature ratio of the sterile neutrino and the background does not change after it becomes non-relativistic.
Finally, by equating $T_{s,NR}=T_{w,NR}$, we can find the relation 
\dis{
\frac{m_s}{T_s} = \frac{m_w}{T_w}.
}
Using the known relations
\dis{
\Omega_w h^2 = \bfrac{T_w}{T_\nu}^3\frac{m_w}{94\ev}, \qquad \frac{T_s}{T_\nu}=\bfrac{10.75}{g_*(T_{NR})}^{1/3},
}
 we obtain the one-to-one correspondence between the lower bound on $m_s$ and the lower bound on $m_w$ as
\begin{equation}
    m_s^{L-\alpha} = 4.46 \ \text{keV} \left(\frac{m_w^{L-\alpha}}{\text{keV}}\right)^{4/3}\left(\frac{10.75}{g_*}\right)^{1/3}\left(\frac{0.12}{\Omega_s h^2}\right)^{1/3},
\end{equation}
where $m_w^{L-\alpha}$ is given in~\eq{Lyman_WDM}.

\section{Production with Dodelson-Widrow mechanism}\label{DW}
In this section, we consider a well-studied model for the production of sterile neutrino called Dodelson-Widrow mechanism~\cite{Dodelson:1993je}, which demonstrates that oscillations between active and sterile neutrinos can yield a sterile neutrino population abundant enough to comprise all or part of the dark matter. In this mechanism, the sterile neutrinos are produced when the active neutrinos are in thermal equilibrium ($T \gg $ MeV). The Boltzmann equation for the evolution of the distribution function of the sterile neutrino $f_s(E,t)$ is given by~\cite{Dodelson:1993je,Abazajian:2001nj}
\begin{equation}
   \frac{\partial f_s(E,t)}{\partial t}-H E \frac{\partial f_s(E,t)}{\partial E}=\frac{1}{4} \sin^2(2\theta_M)  \Gamma_e  [f_e(E,t) - f_s(E,t)]
\end{equation}
where $H= \sqrt{\pi^2 g_*/30} T^2/M_{\text{pl}}$ is Hubble parameter in the radiation dominated Universe, with  reduced Planck mass $M_{\text{pl}} = 2.4 \times 10^{18}\ \text{GeV}$ and $g_*$ the effective degrees of freedom of the relativistic particles in the thermal equilibrium. $f_e(E,t)$ represents the distribution function of electron neutrino and the total interaction rate $\Gamma_e$ between the electron neutrinos and the plasma is given by
\begin{equation}
    \Gamma_e \approx 1.27 \times G_F^2 T^4 E.
\end{equation}
Since our main purpose is to detect sterile neutrinos using the PTOLEMY-like experiments, in this paper, we will only consider electron-sterile mixing. Here, the effective mixing angle in the matter between the sterile and electron neutrino is given by~\cite{Notzold:1987ik,Abazajian:2005gj, Asaka:2006nq}
\begin{equation}
    \sin^2(2\theta_M) = \frac{\sin^2(2\theta)}{\sin^2(2\theta)+[\cos(2\theta)-2 E \:V_T(T)/m_s^2]^2},
    \label{eq:sinM}
\end{equation}
with mixing angle $\theta$ in the vacuum and
\begin{equation}
    V_T =-B T^4E,\quad \textrm{and}\quad  B\sim\begin{cases}
    10.88\times 10^{-9}\ \text{GeV}^{-4} & T>2 m_{e}\\
    3.04\times 10^{-9}\ \text{GeV}^{-4} & T<2 m_{e}
    \end{cases}.
    \label{VT}
\end{equation}
By using relations $y\equiv E/T$ and $t=1/(2H)$ for radiation dominated era, the equation is simplified as
\begin{equation}
    HT\left(\frac{\partial f_{s} (y,T)}{\partial T}\right)_{y\equiv E/T} \simeq -\frac14 \sin^2(2\theta_M) \Gamma_e[f_{e}-f_s],
    \label{eq:fs}
\end{equation}
where the partial derivative about $T$ in the left-hand side is evaluated assuming constant $g_*$. The distribution function of the electron neutrino $f_e$ is assumed to be in the thermal equilibrium and constant with time for fixed $E/T$ as $f_e = (\exp(y)+1)^{-1}$.

 We can find the general solution for $f_{s}$ by redefining $f_s=f_e(1-e^{-f_{s,0}/f_e})$ in \eq{eq:fs} and solving differential equation for $f_{s,0}$ which is given by
\begin{equation}
    HT\left(\frac{\partial f_{s,0} (y,T)}{\partial T}\right)_{y\equiv E/T} \simeq -\frac14 \sin^2(2\theta_M) \Gamma_e f_e.
    \label{eq:fs0}
\end{equation}
The integral solution for $f_{s,0}$ is
\begin{equation}
    f_{s,0}(y,T)\simeq-f_e\int^{T}_{\infty}\frac{1}{4HT}\sin^2(2\theta_M)\Gamma_e dT.
    \label{eq:sol_fs0}
\end{equation}
In fact, $f_{s}$ approaches to $f_{s,0}$ for $f_{s}\ll f_e$, which corresponds to the case when the mixing of the sterile neutrinos is small enough. 
Here, we count $g_*$ of the thermal particles in the standard model and neglect the contribution from the sterile neutrino, which is subdominant. We checked that our result is consistent with that using the program \texttt{LASAGNA}~\cite{Hannestad:2013pha}.

Since the mixing in the matter is suppressed at high temperature, the production rate of the sterile neutrino in the ratio $d(n_s/n_e)/d\log T$ is maximum at a temperature $T_{\text{max}}$~\cite{Dodelson:1993je}
\begin{equation}
    T_{\text{max}}\simeq
    108 \ \text{MeV} \left(\frac{m_s}{\text{keV}}\right)^{1/3},
    \label{Tmax}
\end{equation}
under an assumption of constant $g_*$, where $m_s$ is the mass of the sterile neutrino. 
The numerical coefficient is slightly different from that in~\cite{Dodelson:1993je} as it is sensitive with the number in~\eq{VT}, which is determined by the flavor of the active neutrino.
Therefore, for $T\gg T_{\text{max}}$, the abundance becomes independent of the temperature and
 the relic density of the sterile neutrino at present is given by~\cite{Kusenko:2009up,Dasgupta:2021ies}
 
\begin{equation}\label{eq:sterile_abundance}
    \Omega_s h^2\equiv \frac{n_s m_s}{\rho_c/h^2} \approx 0.1 \left( \frac{\sin^2 \theta}{3 \times 10^{-9}}\right) \left(\frac{m_s}{3 \ \text{keV}}\right)^{1.8},
\end{equation}

where $n_s = \frac{g_sT_0^3}{(2\pi)^3} \int f_{s}(y,T_0) d^3y$ at present temperature $T_0$, the present Hubble parameter $H_0=100 \, h\km/(\sec\mpc)$ and the critical energy density $\rho_c \equiv 3M_P^2 H_0^2\simeq 10^{-5}\gev \cm^{-3}$.

Once the present relic density of the non-relativistic sterile neutrino is given, we can estimate the number density in the early Universe when the sterile neutrinos are relativistic, 
\dis{
n_s(a) = \frac{\Omega_s\rho_c}{m_s}\bfrac{a_0}{a}^3,
}
where $a$ is the scale factor and $a_0$ is its value at present.
Since the number density of the electron neutrino can be obtained in the same way, we can write $\Delta N_{\text{eff}} $ as
\dis{
\Delta N_{\text{eff}} =\left(\frac{\rho_s}{\rho_\nu}\right)_{\rm CMB} \simeq \frac{n_s(a)}{n_\nu(a)}=\frac{\Omega_s h^2 /m_s}{\Omega_\nu h^2/m_\nu} < \Delta N_{\text{eff}}^{\rm max},
\label{DNeff_s}
}
where $\Omega_\nu h^2= \frac{m_\nu}{94 \ {\rm eV}}$ is the energy density of the non-relativistic single flavor neutrino at present.

The local density of the sterile neutrinos near the Earth is enhanced due to the clustering of the gravitational interaction.
Using \eq{nsloc} and \eq{DNeff_s}, the local number density of the sterile neutrino in the DW mechanism is given by
\dis{
    n_{s,{\text{loc}}}
 =& \Delta N_{\text{eff}} [1+f_c(m_s)] n_{\nu},
}
where $n_\nu = 112\ {\text{cm}}^{-3}$ is the global number density of the active neutrino in the present Universe. Therefore, 
the total number of event of $C\nu_SB$ for time $t_{\rm yr}$ becomes 
\begin{equation}\label{capturerate_cluster}
   N_s \simeq  t_{\rm yr} N_T \bar{\sigma}  |U_{e4}|^2 \Delta N_{\text{eff}}  [1+f_c(m_s)] n_{\nu}.
\end{equation}

\begin{figure}[tbp]
\begin{center}
 \includegraphics[width=0.45\textwidth]{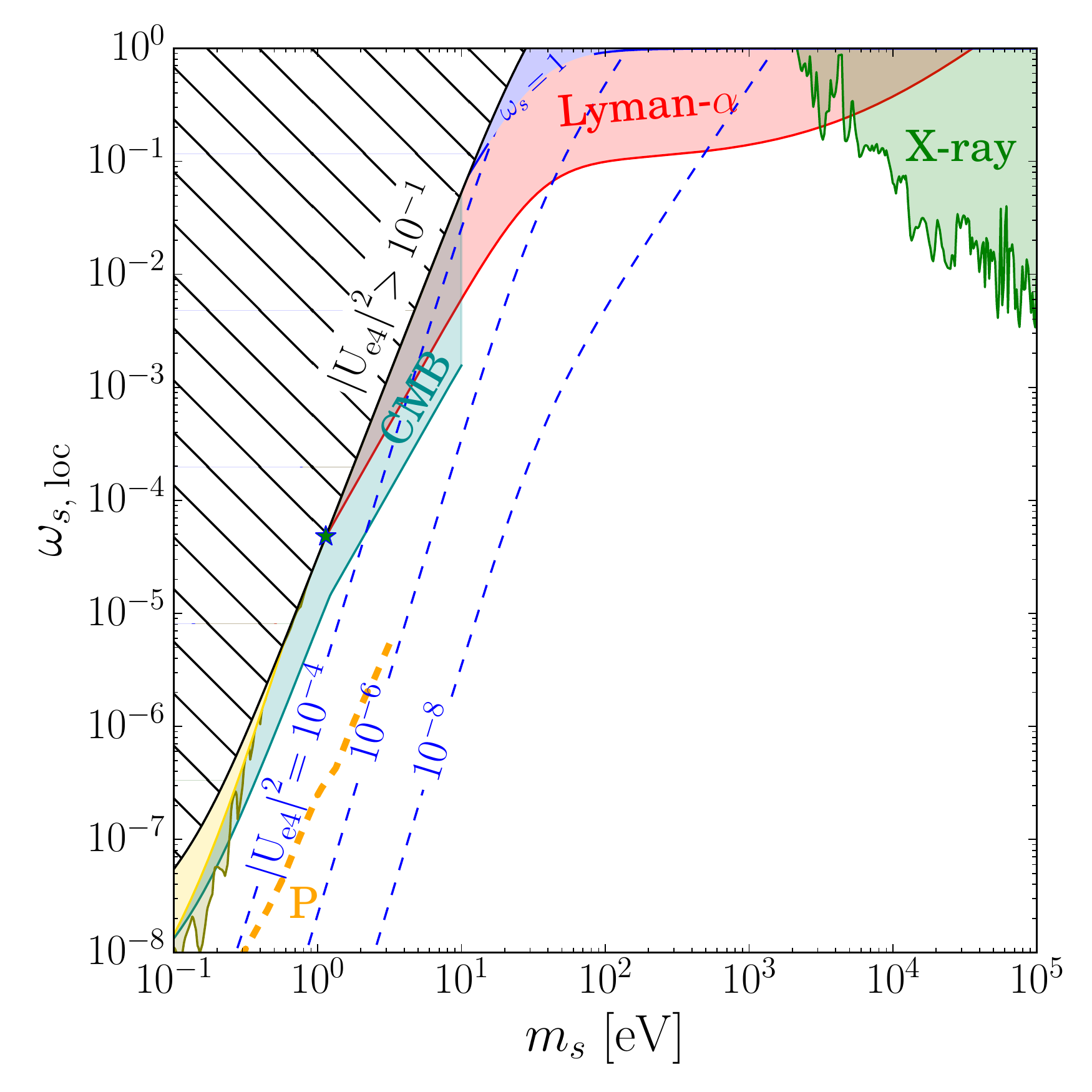}
\hfill
 \includegraphics[width=0.45\textwidth]{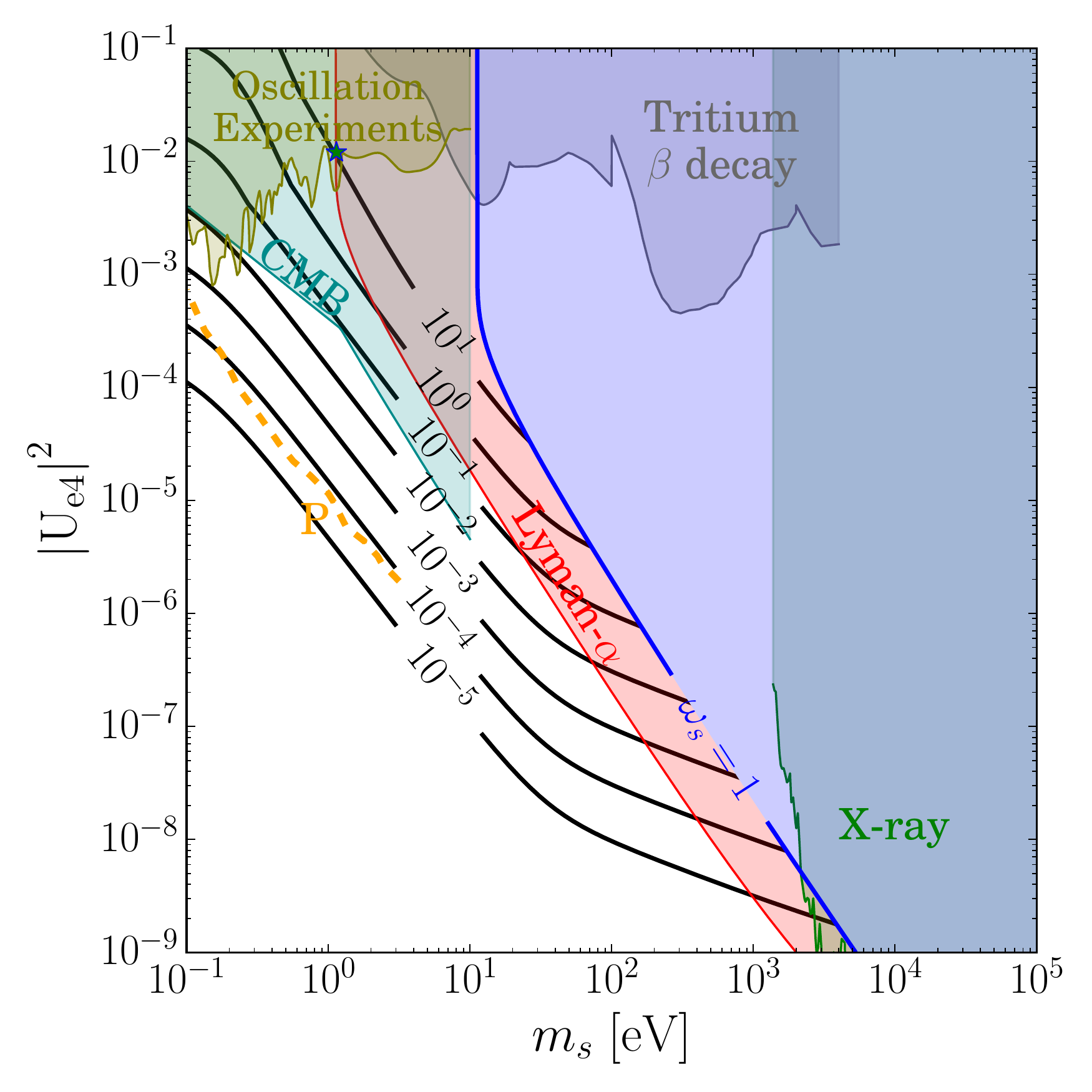}
\end{center}
\caption{{\bf Left:} The fraction of local energy density $\omega_{s,{\rm loc}}$ of sterile neutrino DM in the Dodelson-Widrow model near the Earth with blue dashed lines for corresponding mixing
$\left|U_{e4}\right|^2 = 10^{-8}$, $10^{-6}$, $10^{-4}$.
{\bf Right:} Contour of the number of events of the sterile neutrino DM for one year in the PTOLEMY-like experiments in the Dodelson-Widrow model on the plane of $m_s$ and $\left|U_{e4}\right|^2$. {\bf Constraints:} The constraints described in Section~\ref{constraints} are shown: 
the phase space bound (yellow), the Lyman-$\alpha$ forest bound (red)~\cite{Hooper:2022byl}, the CMB (dark cyan)~\cite{Planck:2018vyg}, and the X-ray (green)~\cite{Hooper:2022byl}, the tritium $\beta$-decay (grey)~\cite{Hiddemann:1995ce,Belesev:2013cba,Kraus:2012he,Abdurashitov:2017kka,KATRIN:2022ith,KATRIN:2022spi}, and DayaBay+Bugey3 experiments (olive)~\cite{DayaBay:2016lkk}. Blue star represents the best-fit point from the short-baseline experiment~\cite{Dentler:2018sju, Gariazzo:2018mwd, Dentler:2017tkw}.
 The orange line shows the expected sensitivity of PTOLEMY by detecting kink and distortion of the $\beta$ decay spectrum~\cite{PTOLEMY:2019hkd}. }
\label{fig:DW}
\end{figure}

In figure~\ref{fig:DW}, we show  {\bf (Left)} the fraction of local energy density $\omega_{s,{\rm loc}}$ of sterile neutrino DM in the Dodelson-Widrow model near the Earth with blue dashed lines for corresponding mixing
$\left|U_{e4}\right|^2 = 10^{-8}$, $10^{-6}$, $10^{-4}$, and 
{\bf (Right)} contour of the total number of events of the sterile neutrino DM for one year with  $100$ g tritium in the PTOLEMY-like experiments  for the Dodelson-Widrow model on the plane of $m_s$ and $\left|U_{e4}\right|^2$.
The constraints described in Section~\ref{constraints} are shown: the phase space bound (yellow),
 the Lyman-$\alpha$ forest bound (red)~\cite{Hooper:2022byl}, the CMB (dark cyan)~\cite{Planck:2018vyg}, the X-ray (green)~\cite{Hooper:2022byl}, the tritium $\beta$-decay (grey)~\cite{Hiddemann:1995ce,Belesev:2013cba,Kraus:2012he,Abdurashitov:2017kka,KATRIN:2022ith,KATRIN:2022spi}, and DayaBay+Bugey3 experiments (olive)~\cite{DayaBay:2016lkk}. Blue star represents the best-fit point from the short-baseline experiment~\cite{Dentler:2018sju, Gariazzo:2018mwd, Dentler:2017tkw}.
 We cut the large mixing $\left|U_{e4}\right|^2>0.1$ with black dashed region in the left figure due to current constraints.
The orange dashed line shows the expected sensitivity of PTOLEMY in the future by detecting kink and distortion of the $\beta$ decay spectrum~\cite{PTOLEMY:2019hkd}. 

For the sterile neutrino produced in the standard Dodelson-Widrow mechanism, the mixing is constrained mostly by the CMB and Lyman-$\alpha$ forest. The most probable number of events is 0.01-0.1 per year for the mass of the sterile neutrino around 1 eV - 100 eV, which is quite difficult to see in the real experiments.
However, this result may change in different production mechanisms of the sterile neutrino, that we will discuss in the next section.

\section{Sterile neutrino DM in the model of low reheating temperature}
\label{lowT}
When the temperature of the early Universe is lower than $T_{\text{max}}$, \eq{eq:sterile_abundance} cannot be applied any more.
In this case, the production of the sterile neutrino is  suppressed and the cosmological and astrophysical constraints can be relaxed~\cite{Gelmini_2004,Bezrukov_2017,Hasegawa:2020ctq}. This can happen when the reheating temperature after inflation is very low or the phase transition for generating the Majorana mass of the sterile neutrino occurs very late.

\subsection{Low reheating temperature}\label{sec:LRT}
When the reheating temperature $T_R$ is smaller than $T_{\text{max}}$, $T_R\ll T_{\text{max}}$, the abundance of the sterile neutrino cannot reach the value in \eq{eq:sterile_abundance}. 
By solving \eq{eq:fs} up to the temperature $T_R\ll T_{\rm max}$, the distribution function of sterile neutrinos can be obtained as
\begin{equation}
     f_s(E,T_0) = \int_{T_R}^{T_0}\frac{\partial f_s}{\partial T} dT\simeq 0.13 \ |U_{e4}|^2 \left(\frac{10.75}{g_*} \right )^{1/2} \left(\frac{T_R}{\text{MeV}} \right)^3  \left(\frac{E}{T_0}\right)f_e(E,T_0),
\end{equation}
where $ |U_{e4}|^2\simeq \sin^2\theta$.
The number density of sterile neutrinos becomes 
\begin{equation}
    n_s \simeq 51.2  \ |U_{e4}|^2 \left(\frac{10.75}{g_*} \right )^{1/2} \left(\frac{T_R}{5 \ \text{MeV}} \right)^3 n_\nu,
\end{equation}
and the relic density is~\cite{Gelmini_2004,Bezrukov_2017}
\begin{equation}
    \Omega_s h^2 \simeq 0.5 \left(\frac{|U_{e4}|^2}{10^{-3}}\right)\left(\frac{10.75}{g_*} \right )^{1/2} \left(\frac{m_s}{\text{keV}} \right) \left(\frac{T_R}{5 \ \text{MeV}} \right)^3.
    \label{Oh2TR}
\end{equation}
Compared to the standard Dodelson-Widrow relic density, a large mixing is needed for small $T_R$ to obtain the given relic density of the sterile neutrino. Accordingly, the cosmological and astrophysical constraints are also relaxed in the $(m_s, |U_{e4}|^2)$ plane to the large mixing. Therefore, the large mixing angle $|U_{e4}|^2\lesssim 10^{-3} $ now survives from the constraints, and a large number of events can be achieved in the future PTOLEMY-like experiment.

\begin{figure}[tbp]
\begin{center}
 \includegraphics[width=0.45\textwidth]{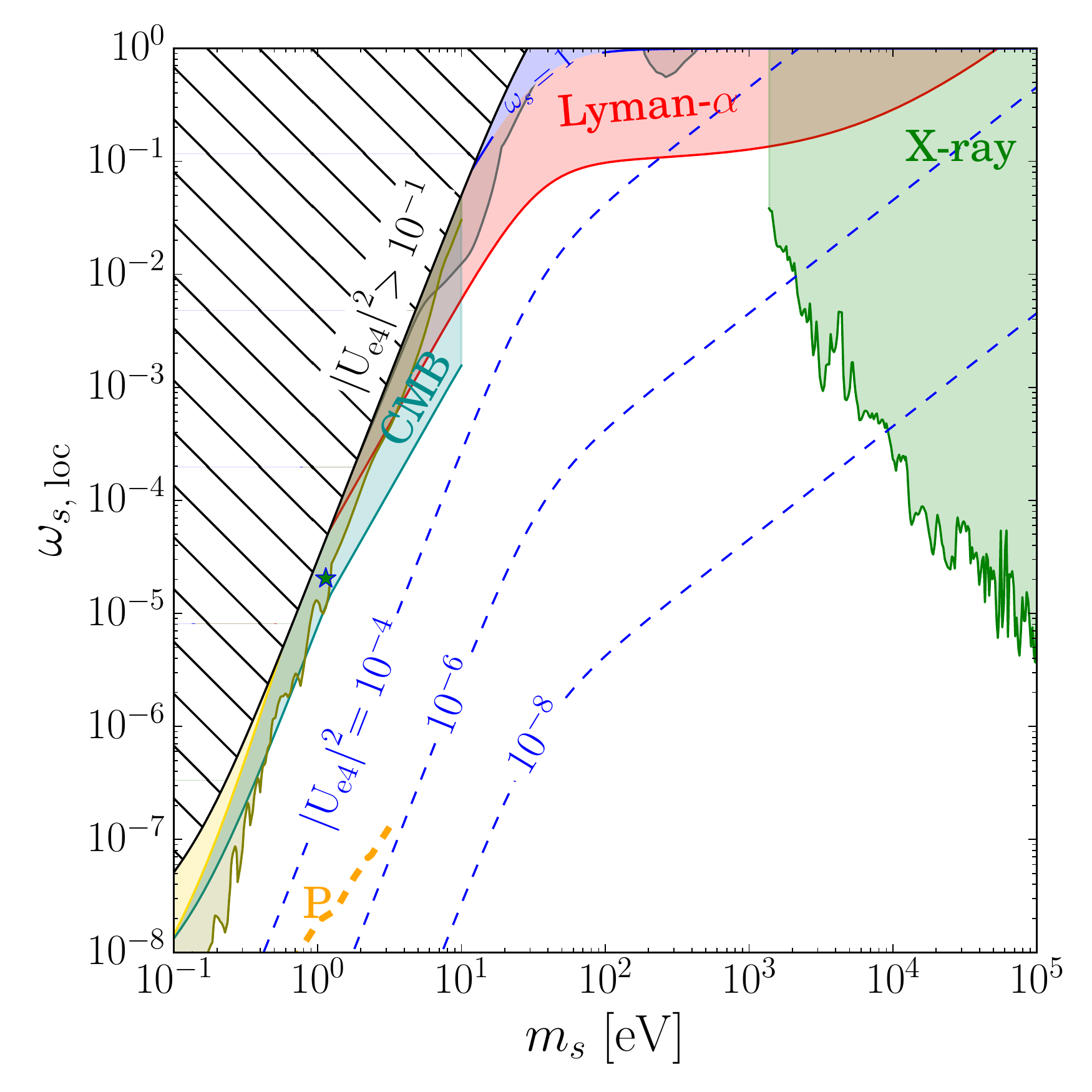}
 \hfill
 \includegraphics[width=0.45\textwidth]{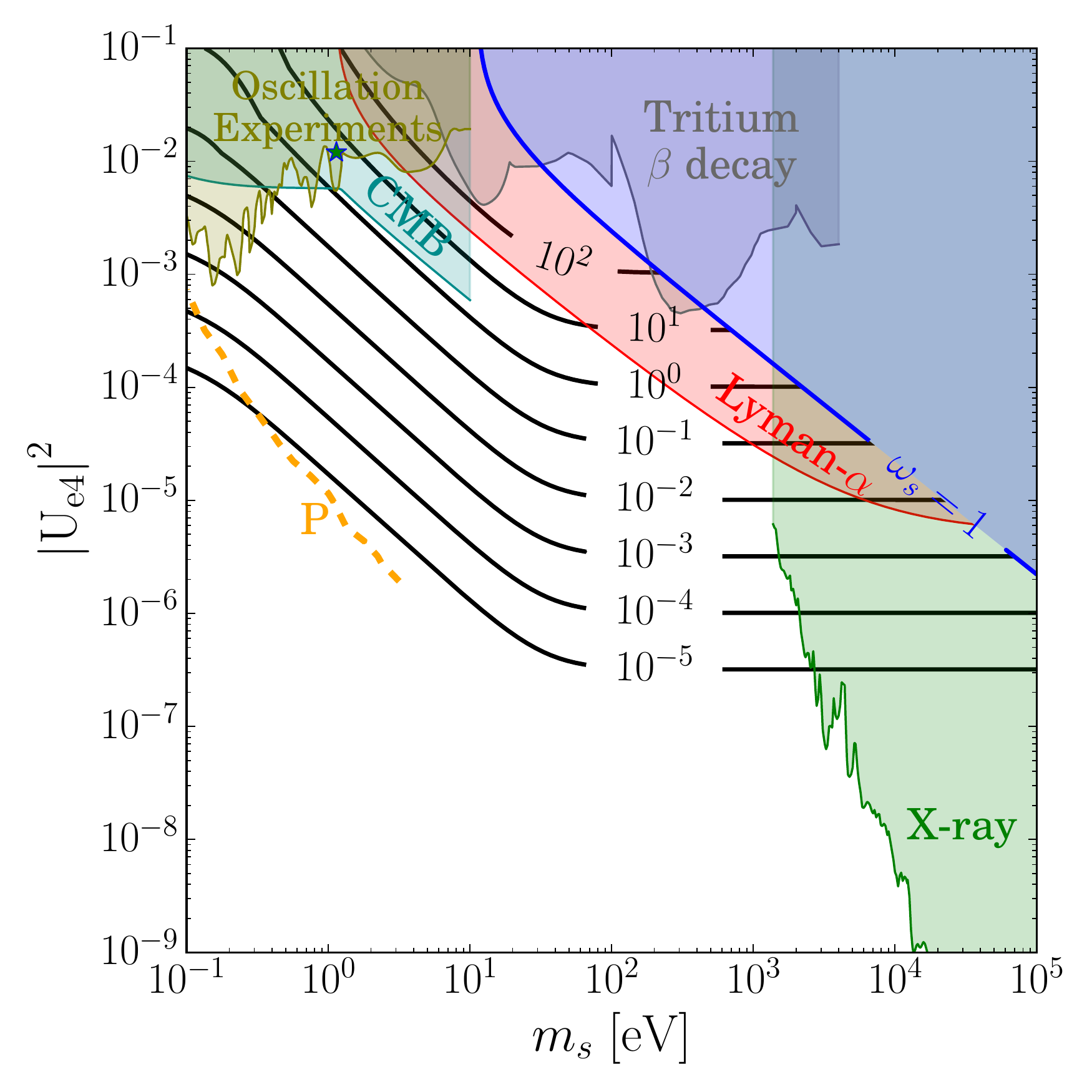}\\
 \includegraphics[width=0.45\textwidth]{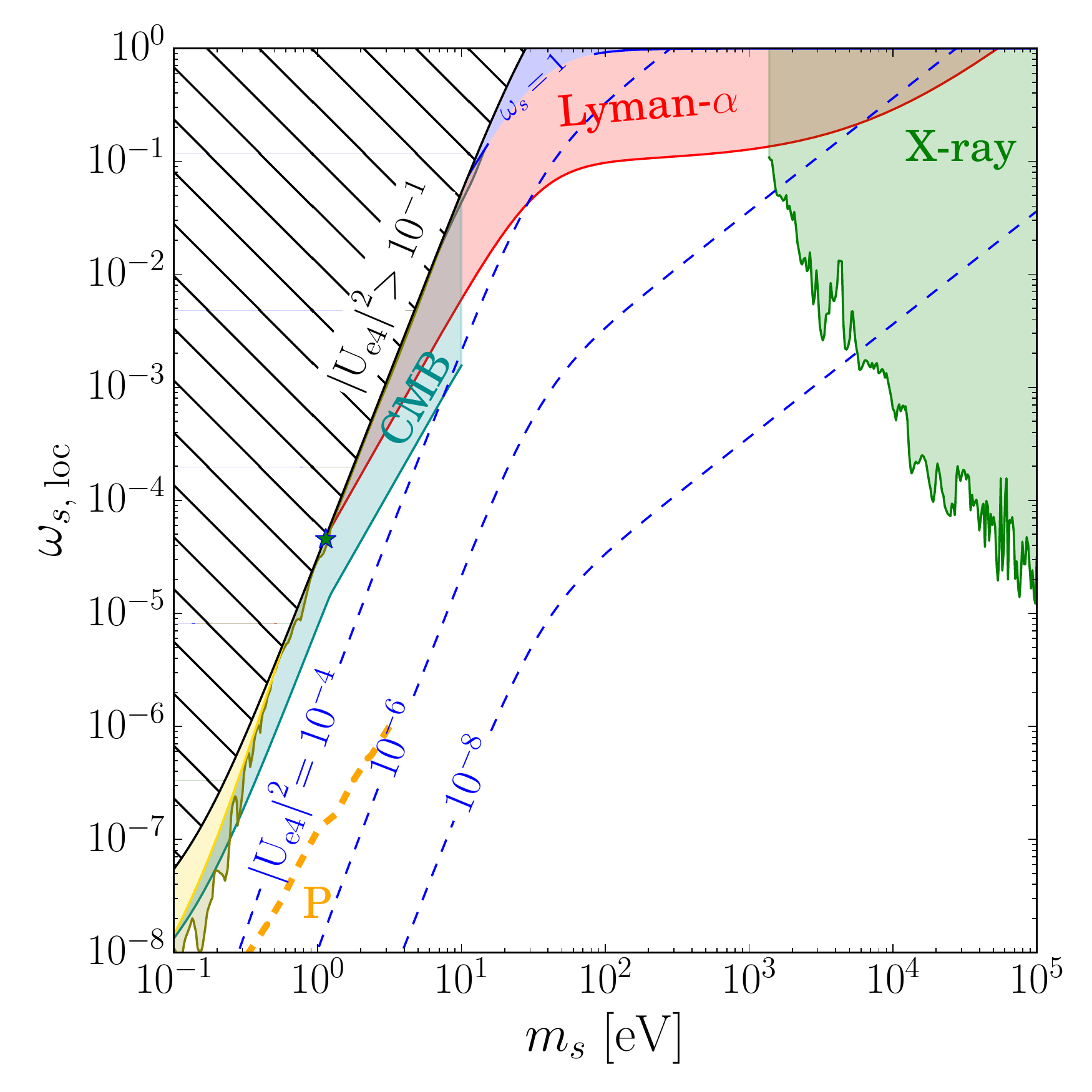}
 \hfill
 \includegraphics[width=0.45\textwidth]{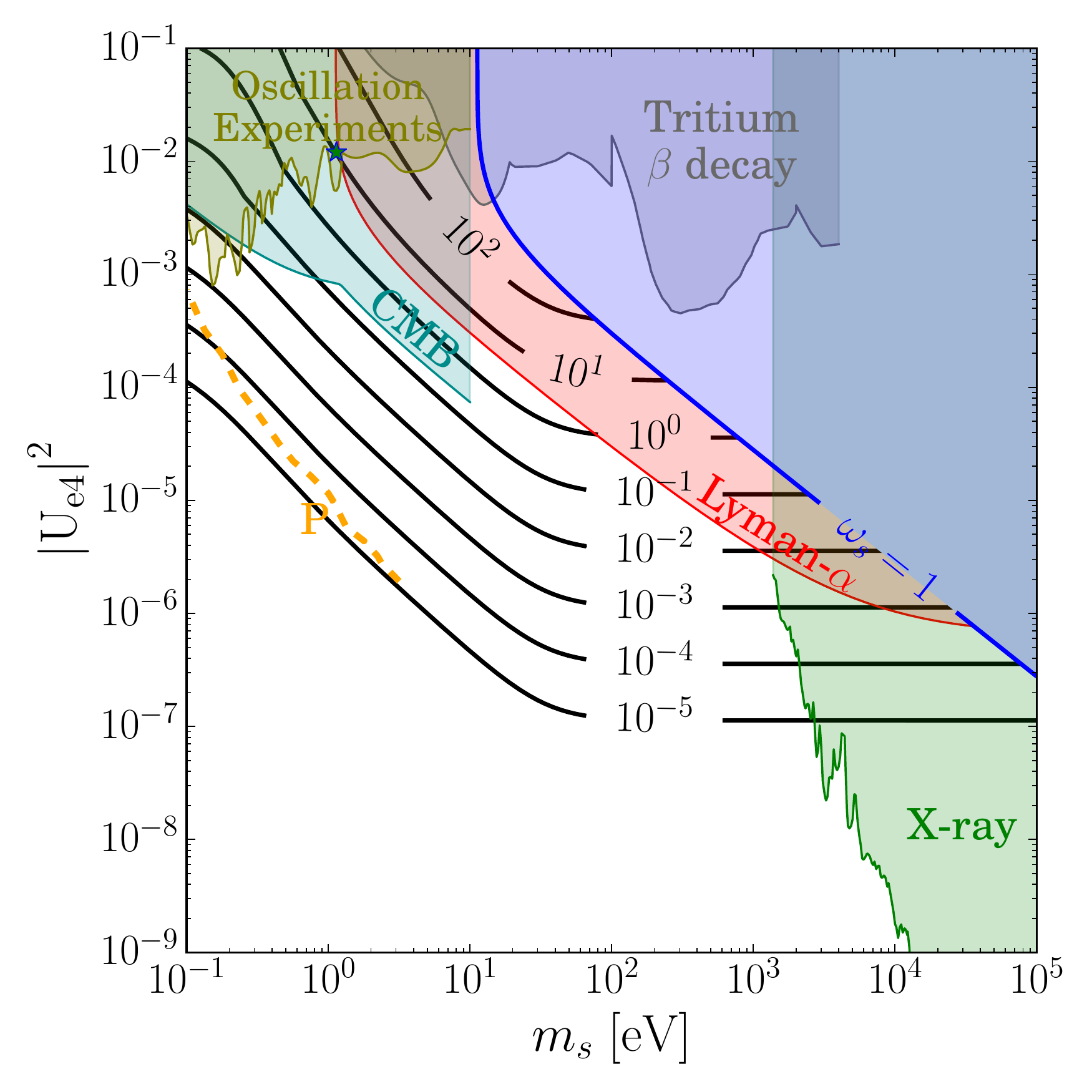}
\end{center}
\caption{Same as figure~\ref{fig:DW} but for the low reheating temperature model with $T_{R}=5\mev{}$ (upper window) and $T_{R}=10\mev{}$ (lower window).}
\label{fig:Reheating}
\end{figure}

In figure~\ref{fig:Reheating}, we show the local DM fraction of the sterile neutrino near the Earth and the number of events in the scenario of the low reheating temperature with $T_{R}=5\mev{}$ (upper window) and $T_{R}=10\mev{}$ (lower window), respectively, to be consistent with BBN~\cite{Hasegawa:2019jsa}. Due to the suppression of the production in the early Universe, the fraction of DM and cosmological constraints appear at large mixing angles, where the capture rate in the $\beta$-decay experiment increases for the same amount of DM fraction in the standard DW mechanism. We find that, for each case, the number of events increases up to ${\mathcal O}(10)$ events, or a few events per year, respectively, at the mass around $10$ eV. The future experiments may probe this model.

\subsection{Late time phase transition in the hidden sector}
In this section, we consider a hidden sector where the phase transition for generating Majorana mass occurs very late after reheating. Before the phase transition, the Majorana mass vanishes and neutrinos comprise Dirac fermion.

We consider a Lagrangian in addition to the standard model~\cite{Bezrukov_2017}
\begin{equation}\label{Lag}
    \mathcal{L} = i \bar{N}\slashed{\partial} N  + Y_\nu H \bar{\nu}_e N_j
    + \frac{\lambda}{2}\phi\bar{N^c} N + h.c., 
    \end{equation}
where $N$ is the right-handed (RH) neutrino with Yukawa interaction with Higgs $H$ and the left-handed (LH) neutrino $\nu$, and also couples to the hidden sector scalar $\phi$ which give Majorana mass to the RH neutrino after the phase transition with $\VEV{\phi}$.
After electro-weak symmetry breaking but before the phase transition, the RH neutrino comprises Dirac fermion with LH neutrino of  mass $M_D = Y_e \VEV{H}$. After the phase transition at temperature $T_c$, the hidden sector scalar $\phi$ develops VEV and gives a Majorana mass $M =\lambda \VEV{\phi}$ to the RH neutrino.

To get the sterile neutrino abundance in this model, we  integrate the Boltzmann equation~\eq{eq:fs} about background temperature $T$, from $T_R$ to the present temperature $T_0$.
For the temperature $T<T_c<T_R$, the sterile neutrino can be produced from the oscillation using the mixing angle in \eq{eq:sinM}. However, for $T_c< T<T_R$, there is no mixing angle since $M=0$, and the sterile neutrino can be generated only through the Dirac mass term.
Therefore the abundance of the sterile neutrino has two contributions
\dis{
\Omega_s = \Omega_{s,c} + \Omega_{s,R},
}
with $\Omega_{s,c} $ from the generation at the temperature between $(T_0,T_c)$ and $\Omega_{s,R}$ between $(T_c,T_R)$.
When $T_c \ll T_{\rm max}$, we can approximate $\Omega_{s,c} h^2$ as in \eq{Oh2TR},
with replacing $T_R$ by $T_c$
\begin{equation}
    \Omega_{s,c} h^2 \simeq 0.5 \left(\frac{|U_{e4}|^2}{10^{-3}}\right)\left(\frac{10.75}{g_*(T_c)} \right )^{1/2} \left(\frac{m_s}{\text{keV}} \right) \left(\frac{T_c}{5 \ \text{MeV}} \right)^3.
    \label{Oh2Tc}
\end{equation}

\begin{figure}[tbp]
\begin{center}
 \includegraphics[width=.45\textwidth]{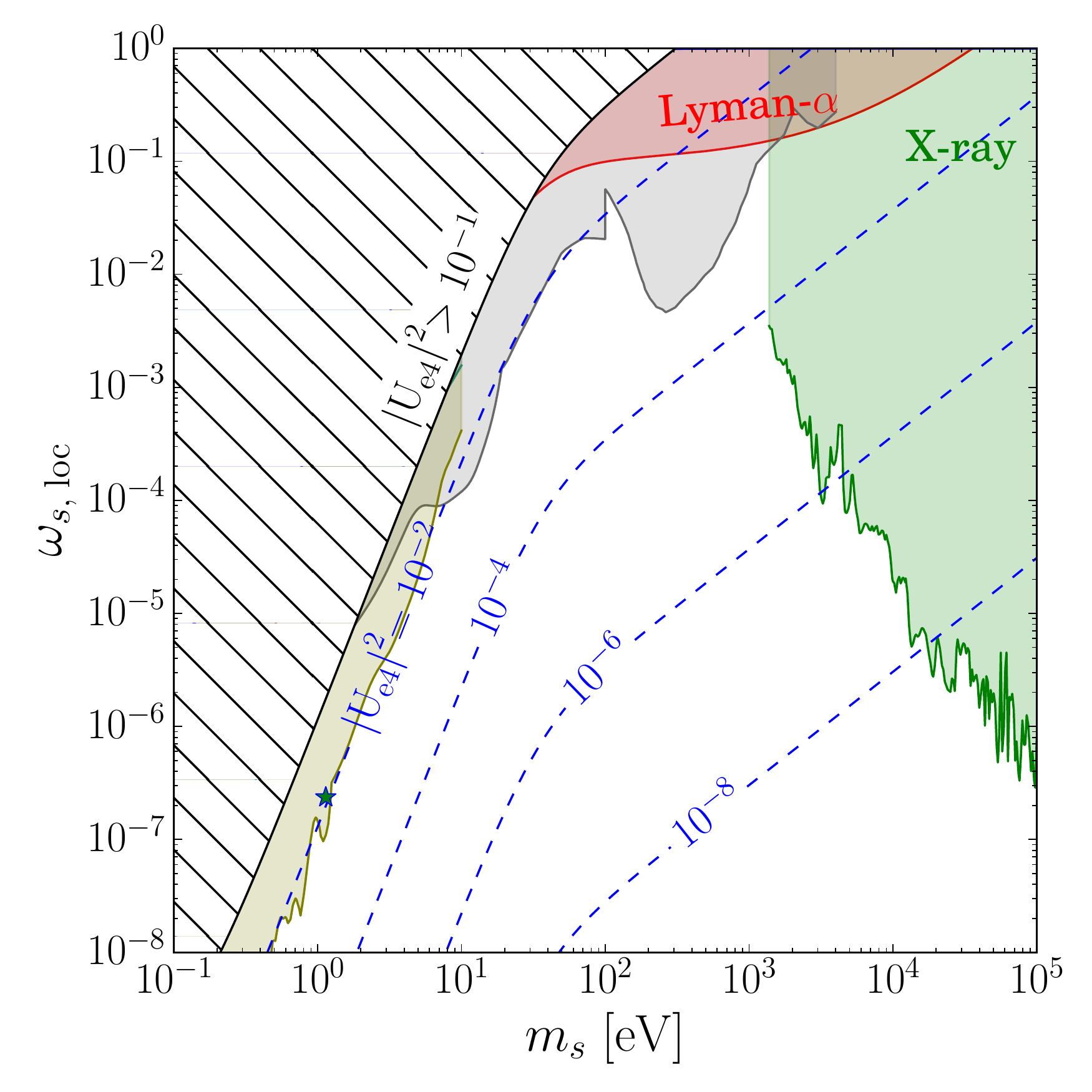}
 \hfill
 \includegraphics[width=.45\textwidth]{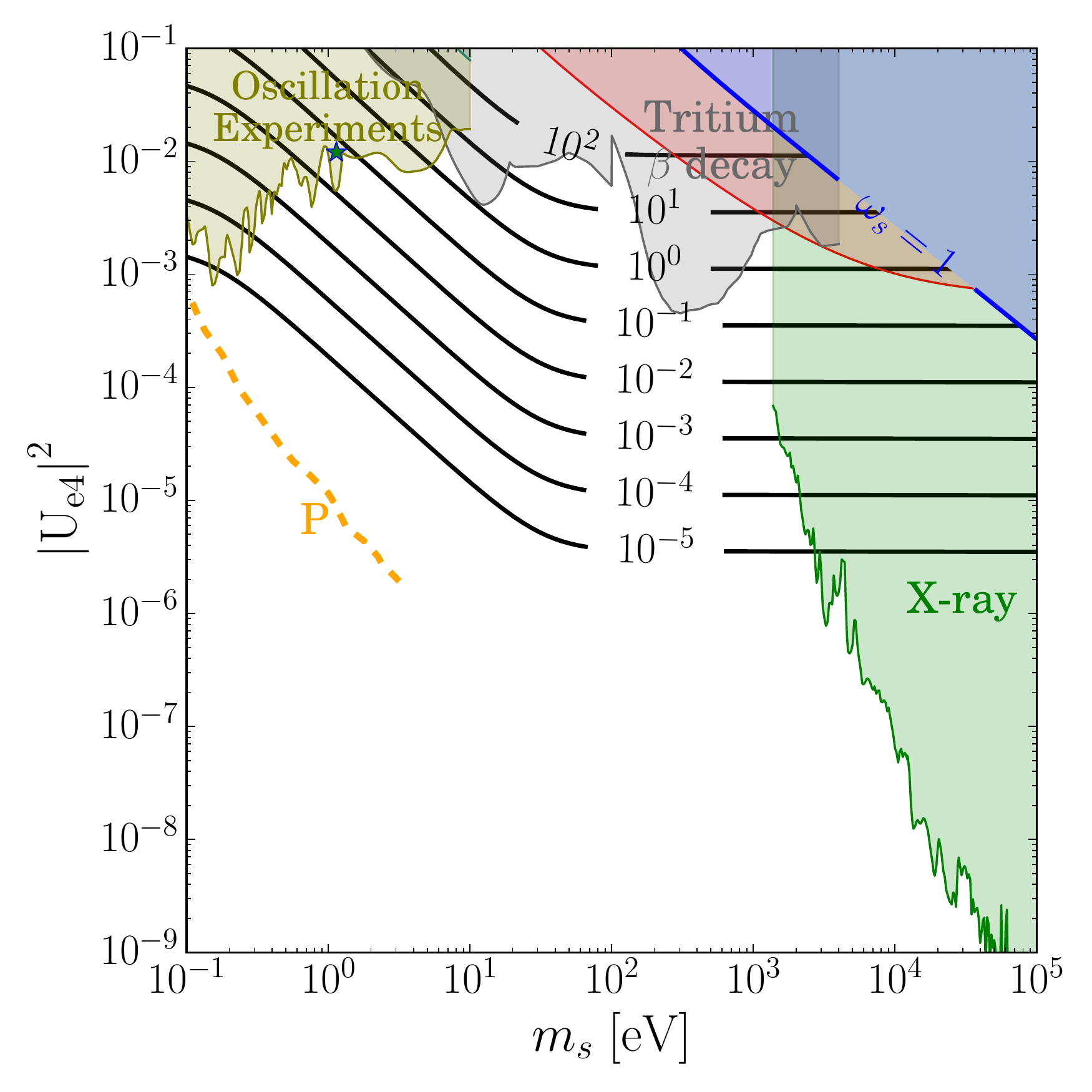}
  \caption{Same as figure~\ref{fig:DW} but for the late phase transition model with $T_{c}=1\mev$ and $T_R=10\mev$.}
\label{fig:PT}
\end{center}
\end{figure}

For temperature $T_c<T<T_R$, there is no mixing term, however sterile neutrino can be produced from the chirality flip in the Dirac mass term which is quite suppressed since it is proportional to $ M_D^2/p^2$. Therefore the Boltzmann equation, \eq{eq:fs} is now modified to 
\begin{equation}
     \left(\frac{\partial f_{s} (E,T)}{\partial T}\right)_{y\equiv E/T} \simeq -\frac{1}{2} \frac{M_D^2}{E^2} \frac{\Gamma_e}{H T} f_{e}, \qquad \Gamma_e \approx 1.27 \times G_F^2 T^4 E,
\end{equation}
where we used $p\simeq E$ for the sterile neutrino since we focus $T_c\gg M$. By integrating this equation, we obtain~\footnote{We find different result from that in~\cite{Bezrukov_2017}. In our case, in the range $T_c<T<T_R$ the amount $f_s/f_e$ is still proportional to $T_R$, however Ref.~\cite{Bezrukov_2017} finds that it is proportional to $T_c$.}
\dis{
    \frac{f_s}{f_e} =& -\int_{T_R}^{T_c}\frac{1}{2} \frac{M_D^2}{E^2} \frac{\Gamma_e}{H T}  dT \approx 0.4 \ \frac{G_F^2 M_{\text{pl}} M_D^2}{\sqrt{g_*(T_R)}} \frac{T_R}{y} \\
    \approx& \frac{8 \times 10^{-7} \ |U_{e4}|^2 }{y}\left(\frac{10.75}{g_*(T_R)}\right)^{1/2}\left(\frac{m_s}{\text{keV}}\right)^2 \left(\frac{T_R}{\text{5 MeV}}\right),
}
where we used $|U_{e4}|^2=M_D^2/m_s^2$.
The number density of sterile neutrino is obtained as
\dis{
  n_{s,R} =& \frac{g}{2\pi^2} \int_{0}^{\infty} dy T^3  y^2 f_s\\
  \simeq& 8 \times 10^{-7} |U_{e4}|^2 \left(\frac{10.75}{g_*}\right)^{1/2}\left(\frac{M_s}{\text{keV}}\right)^2 \left(\frac{T_R}{\text{5 MeV}}\right)   \bfrac{g}{2\pi^2}T^3 \int_0^\infty dy   y f_e,
}
and thus the relic density of sterile neutrino is
\begin{equation}
    \Omega_{s,R} h^2 \approx 4 \times 10^{-9} \ \bfrac{|U_{e4}|^2}{10^{-3}}\left(\frac{10.75}{g_*(T_R)}\right)^{1/2}\left(\frac{m_s}{\text{keV}}\right)^3 \left(\frac{T_R}{\text{5 MeV}}\right).
    \label{OsTcTR}
\end{equation}
The abundance $\Omega_{s,R}$ generated between $T_c<T<T_R$ is quite suppressed and subdominant to $\Omega_{s,c}$ for $T_R\lesssim 10^3\gev$ if $T_c=1\mev$.

In figure~\ref{fig:PT}, we show the numerical result for the fraction of the sterile neutrino and the number of events in this model with $T_c=1 \mev$, and $T_R=10 \mev$. 
Now due to the suppression of the production, larger mixing is needed and the cosmological constraints are hidden behind the constraints from the terrestrial experiments. The maximum number of events around 50 can be available for the sterile neutrino mass $100$ eV and mixing $|U_{e4}|^2\sim 10^{-2}$.

 \section{Conclusion}
 \label{con}
One of the natural ways to explain the neutrino oscillation and the component of dark matter is to introduce right-handed neutrinos. The sterile neutrinos are produced in the early Universe and can be stable enough to survive up to the present time, and comprises a cosmic background as hot, warm, or cold dark matter.
This cosmic neutrino background might be detected with the electron spectrum from the radioactive $\beta$-decaying nuclei, and the prospects of detection strongly depend on the cosmological models and experimental constraints. 

In this paper, we studied the possibility of detecting the cosmic sterile neutrino background in the tritium decay  of future PTOLEMY-like experiments for different models of the sterile neutrino production in the early Universe. We considered two non-standard models with the low-reheating temperature and the late phase transition. In both models, the production of the sterile neutrinos in the early Universe is suppressed and thus it is necessary to have a large mixing between electron and sterile neutrinos. Furthermore, even though the global abundance of the sterile neutrino is smaller  than  the active ones, the gravitational clustering can enhance the local density of the massive sterile neutrinos. 

We find that in both models, the number of events detected can be $\mathcal O(10)$ per year with 100 grams of tritium for the  mass  of sterile neutrino around $10 - 100$ eV and the mixing $|U_{e4}|^2\sim 5\times 10^{-3}$ between electron neutrino and sterile neutrino. In the future PTOLEMY-like experiment, these models can be probed and hopefully the cosmic sterile neutrinos can be detected.

\acknowledgments

The authors were supported by the National Research Foundation of Korea (NRF) grant funded by the Korea government (MEST) (NRF-2022R1A2C1005050).

\bibliographystyle{unsrt}
\bibliography{references}
\end{document}